\documentclass[usenatbib]{mn2e}

\usepackage{aas_macros}
\usepackage{xspace}
\usepackage{psfig}

\newcommand{\um}{\ensuremath{\umu\mathrm{m}}\xspace}
\newcommand{\betapic}{$\beta$~Pic\xspace}
\newcommand{\epseri}{$\epsilon$~Eri\xspace}
\newcommand{\sigboo}{$\sigma$~Boo\xspace}
\newcommand{\iso}{\textit{ISO}\xspace}
\newcommand{\iras}{\textit{IRAS}\xspace}

\title{Sub-mm observations and modelling of Vega type stars}
\author[I.~Sheret, W.~R.~F.~Dent and M.~C.~Wyatt]
{I.~Sheret$^1$\thanks{E-mail: is@roe.ac.uk}, W.~R.~F.~Dent$^2$ and M.~C.~Wyatt$^2$ \\
  $^1$Institute for Astronomy, University of Edinburgh, Royal Observatory, Blackford Hill, Edinburgh, EH9 3HJ \\        
  $^2$UK Astronomy Technology Centre, Royal Observatory, Blackford Hill, Edinburgh, EH9 3HJ}

\date{Accepted 1988 December 15.
      Received 1988 December 14;
      in original form 1988 October 11}

\pagerange{\pageref{firstpage}--\pageref{lastpage}}
\pubyear{2003}

\begin{document}

\maketitle

\label{firstpage}

\begin{abstract}
  We present new sub-mm observations of Vega excess stars, and
  consistent modelling for all known Vega excess stars with sub-mm
  data. Our analysis uses dust grain models with realistic optical
  properties, with the aim of determining physical parameters of the
  unresolved disks from just their SEDs. For the resolved targets, we
  find that different objects require very different dust grain
  properties in order to simultaneously fit the image data and
  SED. Fomalhaut and Vega require solid dust grains, whilst HR4796 and
  HD141569 can only be fitted using porous grains. The older stars
  tend to have less porous grains than younger stars, which may indicate
  that collisions in the disks have reprocessed the initially fluffy
  grains into a more solid form. \epseri appears to be deficient in
  small dust grains compared to our best fitting model. This may show
  that it is important to include all the factors which cause the size
  distribution to depart from a simple power law for grains close to
  the radiation pressure blowout limit. Alternatively, this discrepancy
  may be due to some external influence on the disk (e.g.\ a planet).

  When the model is applied to the unresolved targets, an estimate of
  the disk size can be made. However, the large diversity in dust
  composition for the resolved disks means that we cannot make a
  reliable assumption as to the composition of the grains in an
  unresolved disk, and there is corresponding uncertainty in the disk
  size. In addition, the poor fit for \epseri shows that the model
  cannot always account for the SED even if the disk size is
  known. These two factors mean that it may not be possible to
  determine a disk's size without actually resolving it.
\end{abstract}


\section{Introduction}

In 1983, the \iras satellite detected a large infrared excess from
Vega ($\alpha$ Lyr) during a routine calibration observation
\nocite{1984ApJ...278L..23A}({Aumann} {et~al.} 1984).  Soon, a similar excess was detected from
three more stars (\betapic, Fomalhaut and \epseri). The excess could
only be explained by a shell or ring of dust, heated to between 50 and
125~K by the central star. Quite by accident, \iras had uncovered the
first examples of solid material orbiting a normal main sequence star
other that our own Sun.

The four stars mentioned above have become the prototypes for the
`Vega excess' phenomenon
\nocite{1993prpl.conf.1253B,2000prpl.conf..639G}({Backman} \& {Paresce} 1993; {Lagrange}, {Backman} \&  {Artymowicz} 2000).  A systematic study
of the \iras all-sky survey shows that around 15~per cent of all
main-sequence stars have an IR excess due to circumstellar dust
\nocite{1997Ap&SS.255..103D,1999ApJ...520..215F}({Dominik} \& {The Hjhvega Consortium} 1997; {Fajardo-Acosta} {et~al.} 1999).  Coronographic and
sub-mm imaging has shown the dust is typically in a ring or disk
geometry \nocite{1984Sci...226.1421S,1998Natur.392..788H}({Smith} \& {Terrile} 1984; {Holland} {et~al.} 1998) and that
these systems have much in common with our own solar system, i.e.\ 
solid material in stable coplanar orbits distributed within a region
roughly the size of the Kuiper belt.

Though the Vega phenomenon is common, the poor resolution of current
far-IR and sub-mm instruments means that few Vega excess stars are
close enough to be resolved. For the cases where the disks are
resolved, detailed analysis using realistic models for the optical
properties and size distribution of the dust can be performed. This
type of work has been attempted for \betapic
\nocite{1998A&A...331..291L}({Li} \& {Greenberg} 1998a), HR4796 \nocite{1999A&A...348..557A,
2003ApJ...590..368L}({Augereau} {et~al.} 1999b; {Li} \& {Lunine} 2003a) and Fomalhaut \nocite{2002MNRAS.334..589W}({Wyatt} \& {Dent} 2002). These
models use the observed spatial distribution of dust, and fit the SED
by varying the dust composition and size distribution. However, as
these objects have been analysed on a case-by-case basis, different
models and assumptions are used, which makes it somewhat difficult to
make direct comparisons. In addition, there are so few resolved
targets that it is impossible to look for statistical trends between
disk size, age, stellar mass, etc.  However, there are many unresolved
targets with good SED information, and new instruments such as SIRTF
and SOFIA will produce many more such objects. A method which could
retrieve the physical parameters of a disk from just its SED would be
invaluable. To accomplish this, the detailed properties of the dust
grains in Vega excess disks must be fully understood.

We have tried to tackle these problems in two ways. Firstly, we have
made sub-mm photometric observations of Vega excess stars, to help
constrain their SEDs. Secondly, we have attempted to model all known
Vega excess stars with sub-mm data, using a consistent model with
realistic dust grains.  This allows us to directly compare the results
of modelling for resolved targets, and use the knowledge of dust grain
size and composition gained to estimate the disk size for the
unresolved targets.

\section{Observations}

Sub-mm observations were made using the SCUBA bolometer array on the
James Clerk Maxwell Telescope. The observations were made as part of
several different observing programmes, and hence data was obtained in
different observing modes. In addition, we have reduced unpublished
archive observations of suspected Vega-excess stars to enlarge our
sample.  The observations are summarised in Table~\ref{tab:observations}.

SCUBA can be used in two ways, either for photometry or for mapping.
The distinction arises because the bolometers are spaced too widely to
fully sample the image at the focal plane. Therefore in mapping mode,
the moving secondary mirror is used to add slight offsets to the
telescope pointing, and hence fill in the gaps in the map. Whilst this
is ideal for extended sources, for an observation of an unresolved
object a lot of time is spent where no bolometer can see the source.
To prevent wasting observation time in this way photometry mode can be
used, which does not attempt to make a fully sampled map. In this
mode, the central bolometer is always on source, and hence the
observation will be more sensitive.  The disadvantage is that a fully
sampled map is not produced, so if the object is resolved the measured
flux will underestimate the true flux.  This mode should therefore
only be used if the target is known to be point-like.  Although a full
jiggle map is not used in this mode, a small 9-point map is used to
prevent small pointing errors from affecting the measurement.

For some of the observations an extended photometry mode was used,
which used a 9-point map with a spacing of 5~arcsec between each
point, instead of the normal 2~arcsec spacing. The aim of this was to
ensure that the true flux was measured even if the source was
marginally resolved, whilst retaining most of the sensitivity
advantage of photometry mode. Also, some data was taken using a
12-point pattern, with four points at the centre and eight points
arranged around the central position, at a distance of 7~arcsec. Data
from these extended photometry modes were reduced in exactly the same
way as the conventional photometry.

\begin{table}
\begin{center}
\begin{tabular}{|l|l|l|l|}
\hline
Object   & Date        & Wavelength & Mode \\
         &             & ($\mu$m)  &      \\
\hline
HD17206  & 23 Aug 1997 & 450/850    & Map  \\
         & 27 Aug 1997 & 450/850    & Map  \\
         & 10 Mar 2001 & 450/850    & Ext. Phot.\\
HD23362  &  7 Jun 2000 & 450/850    & Map  \\ 
HD34282  & 25 Sep 1997 & 450/850    & Map  \\
         & 29 Jan 2000 & 450/850    & Map  \\
         & 21 Mar 2000 & 450/850    & Map  \\ 
HD34700  & 14 Aug 1997 & 1350       & Phot \\
         & 13 Dec 1997 & 450/850    & Map  \\
         & 14 Dec 1997 & 450/850    & Map  \\
         & 11 Feb 1998 & 1350       & Phot \\
         & 29 Jan 2000 & 450/850    & Phot \\
HD35187  & 13 Dec 1997 & 450/850    & Map  \\
         & 14 Jan 1998 & 450/850    & Map  \\ 
HD38393  & 11 Mar 2001 & 450/850    & Ext. Phot.\\
         & 12 Mar 2001 & 450/850    & Ext. Phot.\\
HD48682  &  9 Mar 2001 & 450/850    & Ext. Phot.\\
         & 10 Mar 2001 & 450/850    & Ext. Phot.\\
         & 11 Mar 2001 & 450/850    & Ext. Phot.\\
         & 12 Mar 2001 & 450/850    & Ext. Phot.\\  
HD69830  &  9 Mar 2001 & 450/850    & Ext. Phot.\\  
HD81515  & 21 Mar 2000 & 450/850    & Map  \\ 
HD109085 & 21 Mar 2000 & 450/850    & Map \\
         & 10 Mar 2001 & 450/850    & Ext. Phot.\\
         & 11 Mar 2001 & 450/850    & Ext. Phot.\\
         & 12 Mar 2001 & 450/850    & Ext. Phot.\\
HD121617 & 19 Feb 1998 & 1350       & Photometry \\
         & 11 Jun 2000 & 450/850    & Photometry \\
HD123160 & 9  Dec 1997 & 450/850    & Photometry \\
         & 19 Feb 1998 & 450/850    & Photometry \\
         & 10 Jun 2000 & 450/850    & Map \\
HD128167 & 21 Jan 2002 & 450/850    & Ext. Phot.\\
HD139664 & 30 Nov 1997 & 450/850    & Map  \\
         & 13 May 1998 & 450/850    & Map  \\
         & 25 Jun 1998 & 450/850    & Map  \\
         & 11 Jun 2000 & 450/850    & Photometry \\ 
HD141569 & 18 Feb 1998 & 1350       & Photometry \\
         & 19 Feb 1998 & 1350       & Photometry \\
         &  4 May 1998 & 1350       & Photometry \\
         & 13 Jan 1999 & 1350       & Photometry \\
         & 10 Jun 2000 & 450/850    & Photometry \\
         & 11 Jun 2000 & 450/850    & Phot \\ 
HD207129 & 17 May 1998 & 450/850    & Map \& Phot \\
         & 28 May 1998 & 450/850    & Map  \\
\hline
\end{tabular}
\end{center}
\caption{New observations used in our analysis. Observations before
  2000 were extracted from the JCMT archive. The extended photometry
  of HD128167 was observed with a 12~point pattern, all other extended 
  photometry is 9~point. \label{tab:observations}}
\end{table}

\begin{table*}
\begin{center}
\begin{tabular}{llllllll}
\hline
Object   & Other name & Spectral type & Distance & $L_{\mathrm{dust}}/L_{\mathrm{star}}$ & Reference & Estimated age & Reference \\
         &            &               & pc       &                                       &          & Myr           & \\
\hline
HD17206  &            & F5/F6V  & 13.97    &                    &   & & \\
HD23362  &            & K2      & 308.6    & $7.9\times10^{-4}$ & 1 & & \\
HD34282  &            & A0      & 163.9    & 0.39               & 1 & & \\
HD34700  &            & G0      & $>180$   & 0.14               & 2 & & \\
HD35187  &            & A2      & 150      & 0.14               & 1 & & \\
HD38393  &            & F7V     & 8.97     & $5.3\times10^{-6}$ & Best fit model & $1660^{+1580}_{-1391}$ & 8 \\
HD48682  &            & G0V     & 16.4     & $1.2\times10^{-4}$ & Best fit model & \\
HD69830  &            & K0V     & 12.58    & $2.3\times10^{-4}$ & Spline & 600-2000               & 10 \\
HD81515  &            & A5Vm... & 107.0    &                    &   & & \\    
HD109085 &            & F2V     & 18.2     & $4\times10^{-4}$   & Spline & \\
HD121617 &            & A1V     & 180$^a$  & $4.5\times10^{-3}$ & 2 & & \\
HD123160 &            & K5      & $>250$   &                    &   & & \\
HD128167 & \sigboo    & F2V     & 15.5     & $1.8\times10^{-5}$ & Best fit model & $1700^{+1320}_{-720}$  & 8 \\
HD139664 &            & F5IV-V  & 17.5     & $1.9\times10^{-4}$ & Spline & $1120^{+880}_{-875}$   & 8 \\
HD141569 &            & B9      & 99.0     & $8.3\times10^{-3}$ & 2 & $5 \pm 3$              & 7  \\
HD207129 &            & G0V     & 15.6     & $1.0\times10^{-4}$ & Spline & $6030^{+2290}_{-1660}$ & 8 \\
\hline                                                                             
HD22049  & \epseri    & K2V     & 3.22     & $1.1\times10^{-4}$ & Best fit model & $730 \pm 200$          & 10 \\
HD39060  & \betapic   & A5V     & 19.3     & $3\times10^{-3}$   & 3 & $20 \pm 10$            & 6  \\
HD109573 & HR4796     & A0V     & 67       & 0.005              & 4 & $8 \pm 2$              & 9  \\
HD172167 & Vega       & A0V     & 7.76     & $2\times10^{-5}$   & 3 & $354^{+29}_{-87}$      & 5  \\
HD216956 & Fomalhaut  & A3V     & 7.69     & $8\times10^{-5}$   & 3 & $156^{+188}_{-106}$    & 5  \\
\hline
\multicolumn{8}{l}{
\parbox{16.0cm}{
Reference list:
1~\protect{\nocite{1996MNRAS.279..915S}{Sylvester} {et~al.} (1996)};
2~\protect{\nocite{2001MNRAS.327..133S}{Sylvester}, {Dunkin} \&  {Barlow} (2001)};
3~\protect{\nocite{1993prpl.conf.1253B}{Backman} \& {Paresce} (1993)};
4~\protect{\nocite{1991ApJ...383L..79J}{Jura} (1991)}; 
5~\protect{\nocite{2001ApJ...546..352S}{Song} {et~al.} (2001)};
6~\protect{\nocite{1999ApJ...520L.123B}{Barrado y Navascu{\' e}s} {et~al.} (1999)};
7~\protect{\nocite{2000ApJ...544..937W}{Weinberger} {et~al.} (2000)};
8~\protect{\nocite{1999A&A...348..897L}{Lachaume} {et~al.} (1999)};
9~\protect{\nocite{1995ApJ...454..910S}{Stauffer}, {Hartmann} \& {Barrado y  Navascues} (1995)};
10~\protect{\nocite{2000ApJ...533L..41S}{Song} {et~al.} (2000)}. \newline
$^a$No parallax was available, so the distance was estimated by assuming the
absolute visual magnitude for a A1V star is +1.0 \protect{\nocite{1992adps.book.....L}({Lang} 1992, page
144)}, and using the measured visual magnitude of
7.298 from \textsc{simbad}.}}\\
\end{tabular}
\end{center}
\caption{Parameters of stars used in our analysis. The top part of the
table shows objects with new sub-mm data, the bottom part shows
objects where we have analysed published data. Spectral type and
distance are from \textsc{simbad}. \label{tab:starpar}}
\end{table*}

\section{Data reduction}

\subsection{Photometry data}

Photometry data were reduced using the \textsc{surf} package. After
combining the positive and negative beams, the data were flat-fielded
and corrected for atmospheric extinction, estimated using a polynomial
fit to the CSO 225~GHz opacity monitor and the relations in
\nocite{ARCHIBALD_CSOTAU}{Archibald}, {Wagg} \&  {Jenness} (2000). Residual sky emission was estimated using
bolometers well separated from the source, and this signal was
subtracted from the data.  The sky bolometers were those in rings~2
and~3 (30 bolometers in total), and their median was taken as the
average sky signal.  The signal from the central bolometer was then
despiked using a 4~sigma cut.

Calibration sources were observed and reduced in exactly the same way
as the targets.  If a planet was used as a calibrator, the
\textsc{fluxes} package was used to estimate the flux and a correction
was made for the loss of flux because the object was resolved. For the
observations taken with the extended photometry mode, all of the
calibration sources were point-like (i.e.\ secondary calibrators, not
planets). Thus the final photometry is derived under the assumption
that the emission is point-like and centred on the star.  If a target
is extended (i.e.\ marginally resolved) or is offset from the position
of the star, then observations in photometry mode will underestimate
the true flux, by an amount shown in Table~\ref{tab:extphot}. As
expected, the extended photometry modes are less affected than the
conventional photometry mode.

\begin{table}
\begin{center}
\begin{tabular}{lllll}
\hline
Offset & FWHM   & \multicolumn{3}{c}{Relative response} \\
arcsec & arcsec & Normal & Ext.\ 9 pt. & Ext.\ 12 pt.   \\
\hline          
0      & 0      &  1.00  & 1.00        &   1.00         \\
0      & 5      &  0.89  & 0.93        &   0.92         \\
0      & 10     &  0.68  & 0.76        &   0.75         \\
5      & 0      &  0.72  & 0.81        &   0.79         \\
5      & 5      &  0.67  & 0.76        &   0.75         \\
5      & 10     &  0.54  & 0.65        &   0.63         \\
10     & 0      &  0.27  & 0.41        &   0.40         \\
10     & 5      &  0.28  & 0.41        &   0.39         \\
10     & 10     &  0.28  & 0.39        &   0.37         \\
\hline
\end{tabular}
\end{center}
\caption{Comparison of the three photometry modes, showing effect of
  an extended or offset Gaussian source. The table shows the
  ratio of the quoted flux density to the true flux density. The
  PSF FWHM is assumed to be 14~arcsec. \label{tab:extphot}}
\end{table}

\subsection{Maps}
Map data was reduced using custom data reduction software written in
IDL. This system is similar to that developed to reduce data for the
8-mJy extragalactic survey \nocite{2002MNRAS.331..817S}({Scott} {et~al.} 2002). This method is
preferred over the \textsc{surf} software because the residual sky
emission is removed more effectively (a plane is fitted to the
residual emission, rather than a constant), and the `zero footprint'
rebinning method produces a map where each pixel is an independent
measurement, which makes subsequent analysis much simpler.  Source
flux was measured by fitting an elliptical Gaussian to the source. To
ensure that the software was reliable, \textsc{surf} was used to
reduce several sources, and the results from the two reduction systems
were compared to ensure that any differences were well within the
errors.

Calibration was done by calculating the flux conversion factor from
the peak flux of a Gaussian fit to calibration sources. Using the peak
signal rather than the integrated signal means that the measurement is
sensitive to changes in the beam shape \nocite{2001MNRAS.327..697D}({Dunne} \& {Eales} 2001),
but this approach was necessary as most of the calibration sources were
secondary calibrators, and only the peak signal is well determined for
these sources.

\section{Results}\label{sec:results}
The results of the sub-mm observations are shown in
Table~\ref{tab:results}. For objects with multiple observations, the
flux given is the weighted mean of all the data. Of the 15 objects
observed, 7 were not detected, 4 were detected at 850~\um and 450~\um,
and 4 were detected at 850~\um only.

Most of the objects have not been observed in the sub-mm before, but
for objects with existing observations, the new data were checked for
consistency with the old results. HD34282 and HD35187 have both been
observed with UKT14 \nocite{1996MNRAS.279..915S}({Sylvester} {et~al.} 1996): HD34282 had a flux of
$1.3\pm0.3$ Jy at 400~\um and $0.409\pm0.027$ Jy at 800~\um, HD35187
had a flux of $0.115\pm0.022$ Jy at 800~\um. The discrepancy
between these fluxes and the new data is small (2 sigma), however
similar discrepancies have been found in other comparisons of UKT14
and SCUBA data \nocite{2001MNRAS.327..133S}(e.g.\  {Sylvester} {et~al.} 2001). This may be due
to calibration errors in the original data.

For the purposes of this paper, the extended photometry observations
have been reduced in exactly the same way as conventional photometry,
so the result of the data reduction is just the total flux. However, the
data also contains some spatial information, and more complicated
reduction will give some indication of the size of the disk as well as
its flux. This will be tackled in subsequent paper (Wyatt et al., in
preparation).

One of the advantages of SCUBA observations is that they can confirm
the \iras excess is associated with the star, and not simply a
background source (i.e.\ a galaxy or galactic cirrus). Contamination
by background sources is a significant problem, and follow up
observations are important to confirm that the \iras associations are
real. Certainly, not all \iras associations are real, as demonstrated
by observations of 55~Cnc \nocite{2002ApJ...570L..93J}({Jayawardhana} {et~al.} 2002) and HD155826
\nocite{2002ApJ...570..779L}({Lisse} {et~al.} 2002).  SCUBA is a useful tool for confirming
associations because it gives source positions which are significantly
more accurate than those from \iras. \iras detections have a typical
positional uncertainty of about 16~arcsec in the cross-scan direction
and 3~arcsec in the in-scan direction \nocite{1988iras....1.....B}({Beichman} {et~al.} 1988). In
map mode, SCUBA positional uncertainties are about 2~arcsec for bright
sources, limited only by the pointing accuracy of the dish. In
photometry mode, the positional accuracy is determined by the size of
the beam (FWHM of 14~arcsec) so these observations are less useful for
this purpose. From our sample, one object was rejected on the basis of
it's sub-mm position: HD123160 was a strong SCUBA detection, but the
source seemed to be offset from the stellar position by about 10
arcsec. In addition, this star is now thought to be a distant class
III giant \nocite{2002ApJ...567..999K}({Kalas} {et~al.} 2002), and there is evidence of
several galaxies in the optical DSS survey image. This indicates that
the \iras source is probably a background galaxy.

SCUBA is also sensitive to background galaxies, and so there is a
possibility that a background galaxy could fall into the beam of the
observations and cause a spurious detection.  The chances of this
happening depend on the limiting flux of the observations (as there
are many more faint galaxies than bright ones) and the size of the
beam. The extended photometry mode is therefore more susceptible to
this problem, as it is sensitive to sources over $\sim2$ times more
area than conventional photometry. The tentative detection of HD38393
has the highest probability of being due to a background galaxy, as
the source is extremely faint and the observations were taken in
extended photometry mode. This possibility is discussed further in
Section \ref{sec:hd38393}.

The mass of the disks was estimated from the 850~\um flux, using a
dust temperature determined from the peak of the dust emission SED,
and this is shown in Table~\ref{tab:results}. We set the mass
absorption coefficient $\kappa=1.7$ cm$^2$ g$^{-1}$ at 850~\um for
consistency with previous work \nocite{1993ApJ...414..793Z,
  1998Natur.392..788H, 2001MNRAS.327..133S}(e.g.\  {Zuckerman} \& {Becklin} 1993; {Holland} {et~al.} 1998; {Sylvester} {et~al.} 2001). Detailed discussion of
the value of $\kappa$ can be found in \nocite{1994ApJ...421..615P}{Pollack} {et~al.} (1994). It
is important to note that this mass estimate cannot measure the mass
held in grains larger than about 1~mm. This is a key point, as in a
collisional cascade (see Section~\ref{sec:size_dist}), the total mass
of the disk is dominated by the contribution from large bodies, which
contribute little to the 850~\um emission. Our mass estimate only
reflects the amount of mass held in small dust grains, and even then
is model dependent. The dust masses of the new targets are comparable
with those measured in previous studies, ranging from 0.04 lunar
masses (HD38393) to 33~lunar masses (HD141569) for the Vega excess
stars, and hundreds of lunar masses or more for the pre-main-sequence
stars (HD34700, HD35187 and HD34282).

\begin{table*}
\begin{center}
\begin{tabular}{lllll}
\hline
Object       & \multicolumn{2}{c}{Flux (Jy)} & $T_{\mathrm{dust}}$ & Dust mass \\
             & 450$\mu$m & 850$\mu$m & K                   & $M_{\mathrm{moon}}$\\
\hline    
HD17206      & $ < 0.1789 $        & $ < 0.0078 $          &     & $<0.33^a$ \\
HD23362      & $ < 0.17   $        & $ < 0.012 $           &     & $<248^a$  \\
HD34282      & $ 1.925 \pm 0.29 $  & $ 0.384 \pm 0.023 $   &     & $3\times 10^4$ $^b$\\
HD34700      & $ 0.218 \pm 0.037 $ & $ 0.0407 \pm 0.0024 $ &     & $>286^a$  \\
HD35187      & $ 0.268 \pm 0.047 $ & $ 0.061 \pm 0.005 $   &     & $298^a$     \\
HD38393      & $ < 0.0428 $        & $ 0.0024 \pm 0.001 $  & 82  & 0.04      \\
HD48682      & $ < 0.0368 $        & $ 0.0055 \pm 0.0011 $ & 99  & 0.3     \\ 
HD69830      & $ < 2.0248 $        & $ < 0.1005 $          &     & $<3.5^a$  \\
HD81515      & $ < 0.26   $        & $ < 0.02 $            &     & $<49.7^a$ \\
HD109085     & $ < 0.0381 $        & $ 0.0075 \pm 0.0012 $ & 85  & $0.6$     \\  
HD121617     & $ < 0.36   $        & $ < 0.012 $           &     & 85 \\
HD123160$^c$ & $ 0.152 \pm 0.046$  & $ 0.021 \pm 0.007$    &     & \\
HD139664     & $ < 0.39   $        & $ < 0.015 $           &     & $<1.0^a$  \\
HD141569     & $ 0.066 \pm 0.023 $ & $ 0.014 \pm 0.002 $   & 90  & 33      \\        
HD207129     & $ < 1.7 $           & $ < 0.018 $           &     & $<0.95^a$ \\
\sigboo      &                     & $0.0062 \pm 0.0017$         & 62  & 0.5     \\
\hline    
\epseri      & $ 0.225 \pm 0.010 $ & $ 0.040  \pm 0.0015 $ & 85  & 0.1  \\
\betapic     &                     & $ 0.104  \pm 0.010  $ & 103 & 7.8  \\
HR4796       & $ 0.180 \pm 0.15  $ & $ 0.0191 \pm 0.0034 $ & 99  & 19   \\
Vega         &                     & $ 0.0457 \pm 0.0054 $ & 68  & 0.8  \\
Fomalhaut    & $ 0.595 \pm 0.035 $ & $ 0.097  \pm 0.005  $ & 75  & 1.7  \\
\hline   
\multicolumn{5}{l}{$^a$Calculated assuming a dust temperature of 100~K.}\\
\multicolumn{5}{l}{$^b$Taken from \protect{\nocite{2003A&A...398..565P}{Pi{\' e}tu}, {Dutrey} \&  {Kahane} (2003)}.}\\
\multicolumn{5}{l}{$^c$Sub-mm emission is offset by 9 arcsec from the star, and is thought}\\
\multicolumn{5}{l}{\ to be due to a galaxy (see Section~\ref{sec:results}).}
\end{tabular}
\end{center}
\caption{Photometry results and estimated dust masses for our
sample. The top part of the table shows new data, the bottom part
shows existing data. The dust temperature is estimated from the
wavelength of maximum emission for the best fitting SED model.
HD34700 only has a lower limit on its distance, and hence we calculate
a lower limit on its dust mass. These mass estimates are consistent
with existing estimates made for Fomalhaut, Vega, and \betapic in
\protect\nocite{1998Natur.392..788H}{Holland} {et~al.} (1998) and for \epseri in
\protect\nocite{1998ApJ...506L.133G}{Greaves} {et~al.} (1998). \label{tab:results}}
\end{table*}

\section{SED modelling}
Our key objective is to determine the size of a Vega excess disk from
just its SED. Because models have many free parameters, we have tried
to constrain as much as possible using our understanding of the
physical processes taking place in the disks, in particular by setting
the size distribution from theoretical arguments.

The model used here is a modified version of the model developed to
fit the SED and sub-mm image of Fomalhaut \nocite{2002MNRAS.334..589W}({Wyatt} \& {Dent} 2002),
and a more detailed description can be found there. The model is based
on a collisional cascade, where small dust grains are continuously
created by collisions between larger bodies.  The only significant
difference between the model used here and that used in
\nocite{2002MNRAS.334..589W}{Wyatt} \& {Dent} (2002) is that we use a simpler assumption about
the spatial distribution of the dust (see Section~\ref{sec:spatial}
for details).

For this model to be accurate, the disk must be optically thin to
radiation from the central star. If this is not the case, then the
inner parts of the disk will shadow the outer parts from starlight,
and a radiative transfer code would be needed to properly model the
system. We have therefore excluded all the objects where
$L_{\mathrm{dust}}/L_{\mathrm{star}}>0.01$. This ensures that disks
are optically thin, as long as the opening angle is more than
$2^{\circ}$. Measured opening angles are generally larger than this,
e.g.\ $7^{\circ}$ for \betapic \nocite{2000ApJ...539..435H}({Heap} {et~al.} 2000). HD34282,
HD34700, and HD35187 all have
$L_{\mathrm{dust}}/L_{\mathrm{star}}>0.1$, so these objects have not
been included in our analysis.

\begin{table}
\begin{center}
\begin{tabular}{llllll}
\hline
Object  & $p$ & $q_{\mathrm{ice}}$ & $D _{\mathrm{min}}$ & Reduced $\chi^2$ & $R$ \\
        &     &                    & \um                 &                  & AU \\
\hline
Fomalhaut &  0.0  &        &   9.6    &   4.2    &    150          \\ 
          &  0.5  &  0.0   &   18.6   &   14.1   &    150          \\
          &  0.9  &  0.0   &   91.0   &   23.2   &    150          \\
          &  0.5  &  1.0   &   12.5   &   9.3    &    150          \\
HR4796    &  0.0  &        &   10.9   &   16.8   &    70           \\
          &  0.5  &  0.0   &   21.3   &   1.8    &    70           \\
          &  0.9  &  0.0   &   105.2  &   3.2    &    70           \\
          &  0.5  &  1.0   &   14.3   &   25.6   &    70           \\
Vega      &  0.0  &        &   28.3   &   4.2    &    $120\pm13$   \\          
          &  0.5  &  0.0   &   55.9   &   4.9    &    $191\pm16$   \\
          &  0.9  &  0.0   &   281.4  &   4.7    &    $237\pm21$   \\
          &  0.5  &  1.0   &   37.2   &   5.4    &    $92 \pm8$    \\
\epseri   &  0.0  &        &   0.74   &   33.6   &    60           \\
          &  0.5  &  0.0   &   0.44   &   110.9  &    60           \\
          &  0.9  &  0.0   &   0.51   &   139.4  &    60           \\
          &  0.5  &  1.0   &   0.67   &   43.7   &    60           \\
          &  0.0  &        &   3.5    &   2.8    &    60           \\
          &  0.0  &        &   7.0    &   6.6    &    60           \\
          &  0.0  &        &   12.0   &   14.0   &    60           \\
\ ($q=1.9$) &  0.0  &  1.0   &   0.74   &   25.6   &    60           \\
\ ($q=1.8$) &  0.0  &  1.0   &   0.74   &   139.4  &    60           \\
\ ($q=1.76$)&  0.0  &  1.0   &   0.74   &   808.9  &    60           \\
\betapic  &  0.0  &        &   5.9    &   12.2   &    $43.5\pm1.8$ \\          
          &  0.5  &  0.0   &   11.2   &   13.6   &    $62.1\pm2.5$ \\
          &  0.9  &  0.0   &   53.3   &   11.5   &    $82\pm3$     \\
          &  0.5  &  1.0   &   7.6    &   18.6   &    $31.9\pm1.1$ \\
HD141569  &  0.0  &        &   10.3   &   2.0    &    $55.0\pm2.3$ \\          
          &  0.5  &  0.0   &   20.1   &   1.9    &    $84\pm4$     \\
          &  0.9  &  0.0   &   99.5   &   2.4    &    $114\pm5$    \\
          &  0.5  &  1.0   &   13.5   &   3.8    &    $43.8\pm1.7$ \\
HD109085  &  0.0  &        &   4.2    &   4.0    &    $180\pm40$   \\
HD38393   &  0.0  &        &   2.7    &   0.4    &    $200\pm50$   \\
HD48682   &  0.0  &        &   2.3    &   7.0    &    $71\pm15^a$  \\
          &  0.0  &        &   2.3    &   3.1    &    $110\pm21^b$ \\
\sigboo   &  0.0  &        &   3.3    &   2.0    &    $320\pm90$   \\
HD207129  &  0.0  &        &   1.6    &   3.7    &    $260\pm50$   \\
\hline   
\multicolumn{6}{l}{$^a$Assuming secondary does not contribute to \iras fluxes}\\
\multicolumn{6}{l}{$^b$Assuming secondary does contribute to \iras fluxes}\\
\end{tabular}
\end{center}
\caption{Results of the SED fitting, where $p$ is the porosity of the
grains, $q_{\mathrm{ice}}$ is the fraction of this porosity filled
with ice, and $D_{\mathrm{min}}$ is the smallest dust grains in the
system. $D_{\mathrm{min}}$ was determined by the blowout limit due to
radiation pressure, except for \epseri (see
Section~\ref{sec:epseri}). Where no error is quoted on the disk
radius, the size was constrained from existing images and was not a
free parameter of the fits. \label{tab:fits}}
\end{table}

\begin{table*}
\begin{center}
\begin{tabular}{lllll}
\hline
Object    & \multicolumn{4}{c}{Flux (Jy)} \\
          & 12 \um      & 25 \um      & 60 \um      & 100 \um     \\        
\hline
Fomalhaut & $12.1\pm1.2$  & $3.9\pm0.3$   & $10.0\pm0.6$  & $10.7\pm0.6$  \\
HR4796    &               & $4.2\pm0.4$   & $7.1\pm0.9$   & $3.6\pm0.4$   \\
Vega      & $27.8\pm1.4$  & $9.3\pm0.5$   & $9.8\pm0.4$   & $7.6\pm0.4$   \\
\epseri   & $6.6\pm0.3$   & $2.18\pm0.11$ & $1.57\pm0.08$ & $1.99\pm0.22$ \\
\betapic  &               & $10.6\pm0.4$  & $19.1\pm0.8$  & $10.4\pm0.4$  \\
HD38393   & $3.00\pm0.15$ & $0.69\pm0.04$ & $0.19\pm0.03$ & $<0.5$        \\
HD48682   & $1.00\pm0.06$ & $0.29\pm0.03$ & $0.36\pm0.05$ & $<1.0$        \\
HD141569  &               & $2.29\pm0.05$ & $5.3\pm0.8$   & $4.1\pm0.5$   \\
\sigboo   & $1.10\pm0.05$ & $0.32\pm0.02$ & $0.13\pm0.04$ & $<0.5$        \\
HD207129  & $0.60\pm0.03$ & $0.15\pm0.02$ & $0.31\pm0.04$ & $<0.24$       \\
\hline   
\end{tabular}
\end{center}
\caption{Colour corrected \iras fluxes, based on the best fitting
  model. Upper limits are three sigma, and are taken directly from the
  \iras faint source catalogue. HD109085 has been excluded because the best
  fitting model is unlikely to be representative of the true spectral
  shape at any of the \iras wavelengths. \label{tab:iras_fluxes}}
\end{table*}

\subsection{Photometric data}
Mid and far IR photometry data is taken from the \iras faint source
catalogue \nocite{1990BAAS...22Q1325M}({Moshir} {et~al.} 1990). Where available, this is
supplemented by ground based mid-IR data from the literature. Also,
published \iso photometry has been added to help constrain the far-IR
part of the SEDs \nocite{2001A&A...365..545H}(in particular data from {Habing} {et~al.} 2001).

\subsection{Dust grain properties}
Our model of the dust grains is based on the assumption that the
grains in protoplanetary disks are aggregates of interstellar dust
grains, with an additional ice component frozen into them as they
grow.  This type of model was originally developed to simulate
cometary grains \nocite{1982come.coll..131G, 1998A&A...330..375G}({Greenberg} 1982, 1998), but
it has also been successfully applied to the dust grains in accretion
disks \nocite{1994ApJ...421..615P}({Pollack} {et~al.} 1994), and to the debris disks of \betapic
\nocite{1998A&A...331..291L}({Li} \& {Greenberg} 1998a), HR4796 \nocite{1999A&A...348..557A,
2003ApJ...590..368L}({Augereau} {et~al.} 1999b; {Li} \& {Lunine} 2003a), HD141569 \nocite{2003ApJ...594..987L}({Li} \& {Lunine} 2003b) and
HD207129 \nocite{1999A&A...350..875J}({Jourdain de Muizon} {et~al.} 1999).

For the composition of the interstellar grains, we use the core-mantle
model developed in \nocite{1997A&A...323..566L}{Li} \& {Greenberg} (1997), which successfully
fits the observed interstellar extinction and polarisation. This model
has a silicate core surrounded by a UV-processed organic refractory
mantle. We force the silicate/organic-refractory volume ratio to be
1:2, as inferred for cometary grains \nocite{1998A&A...330..375G}({Greenberg} 1998). The
grains are assumed to be porous, with the porosity being a free
parameter of the model. For dust grains cooler than $\sim 120$ K, the
gaps due to porosity may be filled with either vacuum, water ice or a
mixture of the two; for dust grains warmer than $\sim 120$ K most ices
would sublime, so the gaps must be filled with vacuum. The free
parameters in the dust composition model are therefore the porosity
$p$ of the grains and the ice fraction $q_{\mathrm{ice}}$, with the
constraint that for warm grains the ice fraction must be zero.

Optical constants for the materials are taken from
\nocite{1997A&A...323..566L, 1998A&A...331..291L}{Li} \& {Greenberg} (1997, 1998a). The optical
constants for the composite dust grain material are then calculated
using Maxwell-Garnet effective medium theory. Then, the absorption
efficiencies for the dust grains are calculated using Mie theory,
Rayleigh-Gans theory, or geometric optics, depending on the size of the dust
grain. Details of Maxwell-Garnet effective medium theory, and the
methods used to calculate absorption efficiencies are discussed in
\nocite{1983asls.book.....B}{Bohren} \& {Huffman} (1983).

\subsection{Size distribution \label{sec:size_dist}}
The size distribution that results from an infinite collisional
cascade is \nocite{1969JGR....74.2531D}({Dohnanyi} 1969):
\begin{equation}
\label{eqn:size_dist}
n(D) \propto D^{2-3q}
\end{equation}
where $n(D) d D$ is the number of planetesimals of size between $D$
and $D+ d D$, and $q=11/6$. However, this distribution will only hold
for large dust grains, as radiation pressure will blow out small
grains (e.g.\ $<1$ \um) very rapidly, and Poynting-Robertson drag will
cause intermediate sized grains (e.g.\ $<10$ \um) to spiral inwards
toward the star. Which of these two effects is dominant depends on
the optical depth of the dust disk, but for all but the most tenuous
disks the collisional timescale is shorter than the PR timescale, so
PR drag can be safely ignored \nocite{1999ApJ...527..918W}({Wyatt} {et~al.} 1999). Given this,
we assume that the above size distribution holds down to the radiation
blowout limit, when $\beta > 0.5$ (where $\beta=F_{\mathrm{rad}}/F_{\mathrm{grav}}$), at
which point there is a sharp cutoff.

\subsection{Spatial distribution \label{sec:spatial}}
For the spatial distribution of the dust grains, we assume a
infinitely thin, flat ring, with the ring's radius $R$ as a free
parameter. This approach is motivated by the sub-mm images of
Fomalhaut and \epseri, which both show a narrow, well defined ring of
emitting material. This spatial distribution is different from that used
in \nocite{2002MNRAS.334..589W}{Wyatt} \& {Dent} (2002), as here no attempt is made to account
for the eccentricities of the dust grains, which broadens the ring by
about 50~AU. The advantage of the simpler model is that it reduces the
number of free parameters, but should still give a reasonable estimate
of the overall extent of the disk. The Fomalhaut results from the two
versions of the model are almost identical (see
Section~\ref{sec:fomalhaut}), which gives us confidence that this
simplification will not affect our conclusions.

\subsection{Stellar parameters}
Accurate stellar parameters are needed for several reasons. Firstly,
the photospheric flux must be estimated in each band, so that the
excess flux due to dust can be determined. This is particularly
important in the \iras 12~and 25~\um bands, where the photosphere can
be the dominant source of flux. Secondly, the temperature of the dust
depends on the luminosity of the star, and on the shape of the star's
spectrum in the region where most power is emitted (i.e.
UV---optical---near-IR).  The luminosity and stellar spectrum also determine
the magnitude of the radiation forces which act on the dust grains.
Finally, an estimate of the mass of the star is needed to calculate the
gravitational forces.

To determine the stellar parameters, we first obtained the spectral
type of each star from \textsc{simbad}. This was used to estimate the
star's effective temperature, using the calibration provided in
\nocite{1994AJ....107..742G}{Gray} \& {Corbally} (1994). We then extracted a Kurucz model
atmosphere with this effective temperature, assuming solar metallicity
and surface gravity $\log(g)=4.3$ appropriate for dwarf stars. The
model atmosphere was then normalised to fit the K band magnitude. K
band was preferred over optical photometry because it is less
susceptible to the effects of reddening. The luminosity of the star was
then calculated by integrating the normalised model atmosphere.
Finally, the mass of the star was estimated using the observed
spectral type and the table provided by \nocite{1992adps.book.....L}{Lang} (1992, page
132).

\subsection{Modelling method}
A model SED is specified by the grain composition (porosity $p$, ice
fraction $q_{\mathrm{ice}}$), radius of the dust ring ($R$), and the
total dust luminosity. In the cases where the size of the dust ring is
known, the ring radius is fixed to the observed size; for the
unresolved objects the ring radius is left as a free parameter. The
approach we take is to choose a dust composition, and then find the
best fit to the data by varying the remaining free parameters.

The emission spectrum of an individual dust grain depends on its
temperature and on the emission efficiency as a function of
wavelength. The SED of a disk is simply the combined emission of all
the dust grains. For a given dust composition model, the absorption
efficiency is calculated as a function of grain size.  Given this, the
temperature of the dust grains as a function of grain size and
distance from the star can be determined. Then, the total flux as a
function of wavelength is found by integrating over all grain sizes,
weighting with the size distribution given in Equation
\ref{eqn:size_dist}.

The model spectrum is then converted into an SED that can be directly
compared to the observed points to obtain a $\chi^2$ value. To do
this, the spectral response of each filter-instrument combination is
multiplied by the model spectrum, to estimate the broad-band flux that
would be recovered from a real observation of the model spectrum. This
step is particularly important for the \iras data points, which have a
very wide band-pass. By converting the model spectrum into broad-band
fluxes, we avoid having to make colour-correction to the original
data, which would otherwise require detailed knowledge of the true
spectrum of the source.  However, when showing the result of the fit,
a colour-correction must be applied so that the data points can be
plotted in real units. For this purpose, the best fitting model is
used to estimate the colour correction for each point. If the best
model is more than 3~sigma away from a data point, then no
colour-correction is applied to that point, as the model is not likely
to give a good estimate of the true colour-correction.

For some objects it is impossible to fit all the data points
simultaneously, because there is dust at a range of distances from the
star. In these cases, only the long wavelength points are fitted, in
order to obtain an estimate of the overall extent of the ring.

\section{Modelling results}

\subsection{Fomalhaut} \label{sec:fomalhaut}

\begin{figure}
  \begin{center}
    \psfig{figure=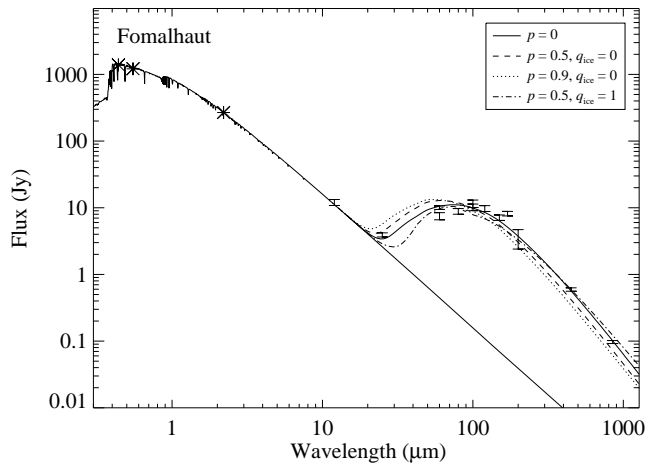,angle=90,width=\columnwidth}
  \end{center}
  \caption{SED of Fomalhaut and model fits with the ring radius set at
    150~AU. Asterisks show optical and near-IR photometry, \iras, \iso
    and SCUBA data are shown with error bars. Colour corrections for
    the \iras data were estimated from the $p=0$
    model. \label{fig:fom_sed}}
\end{figure}

The size of Fomalhaut's dust disk is known from the sub-mm images made
with SCUBA \nocite{2003ApJ...582.1141H}({Holland} {et~al.} 2003). Because the ring is inclined
and the images have low resolution the size of the ring cannot be
measured directly from the images, but the best estimate of the true
size is 150~AU \nocite{2002MNRAS.334..589W}({Wyatt} \& {Dent} 2002). The SED data for Fomalhaut
comes from \iras, SCUBA and also \iso. The \iso fluxes are from Walker et
al. (in preparation), and are listed in \nocite{2002MNRAS.334..589W}{Wyatt} \& {Dent} (2002).

Our modelling (Figure~\ref{fig:fom_sed}) shows a good fit for solid dust grains (i.e.
$p=0.0$), but rules out porous grains as this would make the dust
hotter than is observed. The $p=0.5$, $q_{\mathrm{ice}}=1.0$ model is
also ruled out, as it fails to fit the mid-IR part of the SED.

The reduced $\chi^2$ for the best fitting model is 4.2, which
indicates that the model is not consistent with the data. However, by
far the dominant contribution to $\chi^2$ comes from the \iso data,
which show considerable scatter from any smooth model. The \iso filters
all have broad band passes ($\Delta \lambda / \lambda \sim 0.5$), so
the apparent scatter in the \iso points is unlikely to be due to
structure in the spectrum. A more likely explanation is that
uncertainties on the \iso points have been underestimated, and this
would explain the high reduced $\chi^2$.

Fomalhaut has already been modelled in detail by
\nocite{2002MNRAS.334..589W}{Wyatt} \& {Dent} (2002), and the results here are almost identical
to those results. As the modelling here uses the same data and dust
grain model, it is hardly surprising that we recover the same results.
However, this does confirm that the simpler spatial distribution used
here does not significantly affect the results of the modelling.

\subsection{HR4796} \label{sec:hr4796}

\begin{figure}
  \begin{center}
    \psfig{figure=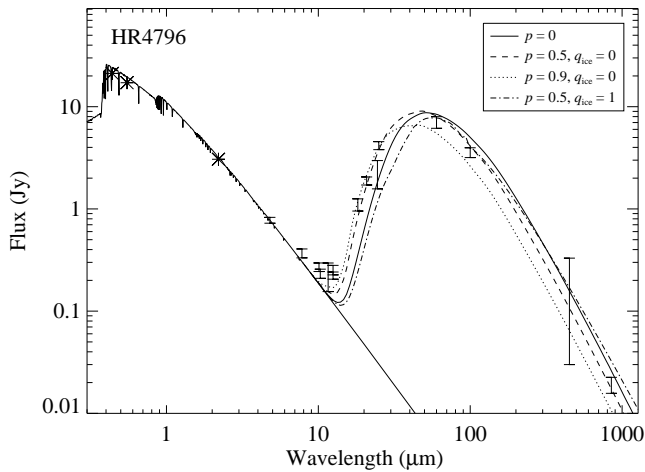,angle=90,width=\columnwidth}
  \end{center}
  \caption{SED of HR4796 and model fits with the ring radius set at 70
    AU. Data where the wavelength is shorter than 18~\um has been
    excluded from the fit, as the excess here is probably due to a hot
    dust component close to the star (see Section~\ref{sec:hr4796}).
    Upper limits are 3~sigma. Colour corrections for the \iras data
    were estimated from the $p=0.5$, $q_{\mathrm{ice}}=0$ model.
    \label{fig:hr4796_sed}}
\end{figure}

HR4796 is resolved in both mid-IR images of dust emission
\nocite{1998ApJ...503L..79J,1998ApJ...503L..83K}({Jayawardhana} {et~al.} 1998; {Koerner} {et~al.} 1998) and scattered light
images in the near-IR \nocite{1999ApJ...513L.127S}({Schneider} {et~al.} 1999). The best estimate
of the size of the ring comes from mid-IR images, which suggest that
the radius is 70~AU. The star is younger than most of the other stars
in the sample, with a age of $8\pm2$ Myr \nocite{1995ApJ...454..910S}({Stauffer} {et~al.} 1995).

The modelling results show that solid dust grains are unable to fit
the observed SED and disk size (Figure~\ref{fig:hr4796_sed}). However,
using a porosity of $p=0.5$ provides a good fit to the SED. The icy
model ($p=0.5$, $q_{\mathrm{ice}}=1.0$) is a worse fit than the solid
grain model, so can be rejected. Around 10~\um, there is an additional
excess which cannot be accounted for by any model. This is likely to
be produced by a hot dust component close to the star. Data in this
part of the spectrum have therefore been excluded from the fit, and do
not contribute to the total $\chi^2$ quoted in Table~\ref{tab:fits}.

Detailed modelling of this object has been done previously by
\nocite{1999A&A...348..557A}{Augereau} {et~al.} (1999b), and our results here are in good
agreement with this work. Augereau et al.  made a much more thorough
exploration of parameter space, but their overall conclusion was that
two dust populations are required: a cool component at 70~AU, and a
hotter component at about 10~AU. Their best fitting model for the cool
dust has amorphous grains with a porosity of $p=0.6$ and an ice
fraction ice $q_{\mathrm{ice}}=0.03$. In their analysis, the ring's
diameter was constrained to be 70~AU, but the grain porosity, ice
fraction and the width of the dust ring were all left as free
parameters. As for Fomalhaut, our results confirm that the simpler
model used here is able to produce similar results to a more detailed
model.

Modelling of HR4796 has also been done by \nocite{2003ApJ...590..368L}{Li} \& {Lunine} (2003a),
and they were able to fit the entire SED without having a hot dust
component, in contrast to the results of
\nocite{1999A&A...348..557A}{Augereau} {et~al.} (1999b) and the results presented here. This
discrepancy is probably because Li and Lunine's model contains a large
number of grains which are much smaller than the blowout diameter, and
these small grains produce a hot component in the SED even though they
are distant from the central star.

\subsection{\epseri \label{sec:epseri}}

\begin{figure}
  \begin{center}
    \psfig{figure=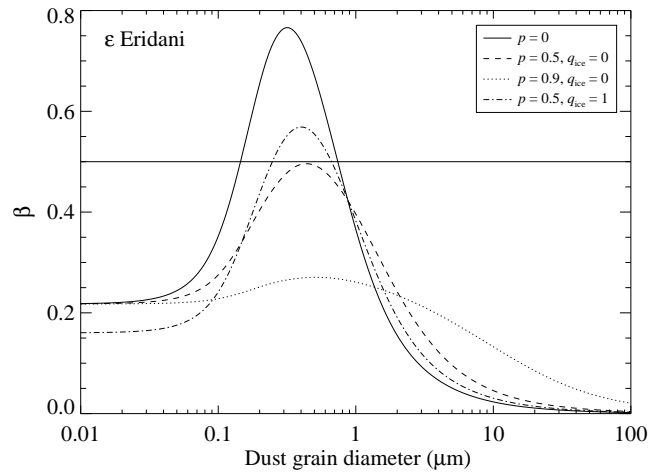,angle=90,width=\columnwidth}
  \end{center}
  \caption{Graph showing how $\beta$
    ($=F_{\mathrm{rad}}/F_{\mathrm{grav}}$) depends on dust grain
    size, for dust orbiting \epseri. $\beta=0.5$ is shown by the
    horizontal line; any dust grain with $\beta > 0.5$ will leave the
    system on a hyperbolic trajectory. \label{fig:epseri_beta}}
\end{figure}

\epseri is a particularly interesting object, as it is the only cool star
(K2V) with a resolved dust disk, all other resolved disks being around
A stars. SCUBA images of \epseri show a distinct ring, which is
approximately face on and has a radius of 60~AU
\nocite{1998ApJ...506L.133G}({Greaves} {et~al.} 1998). 

\begin{figure}
  \begin{center}
    \psfig{figure=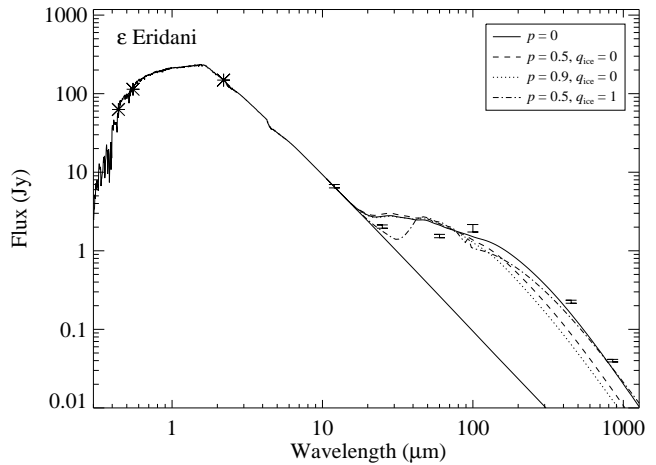,angle=90,width=\columnwidth}
  \end{center}
  \caption{SED of \epseri, with the ring radius set at 60
    AU. Colour corrections for the \iras data as plotted here were
    estimated from the $p=0$ model. \label{fig:epseri_comp}}
\end{figure}

\begin{figure}
  \begin{center}
    \psfig{figure=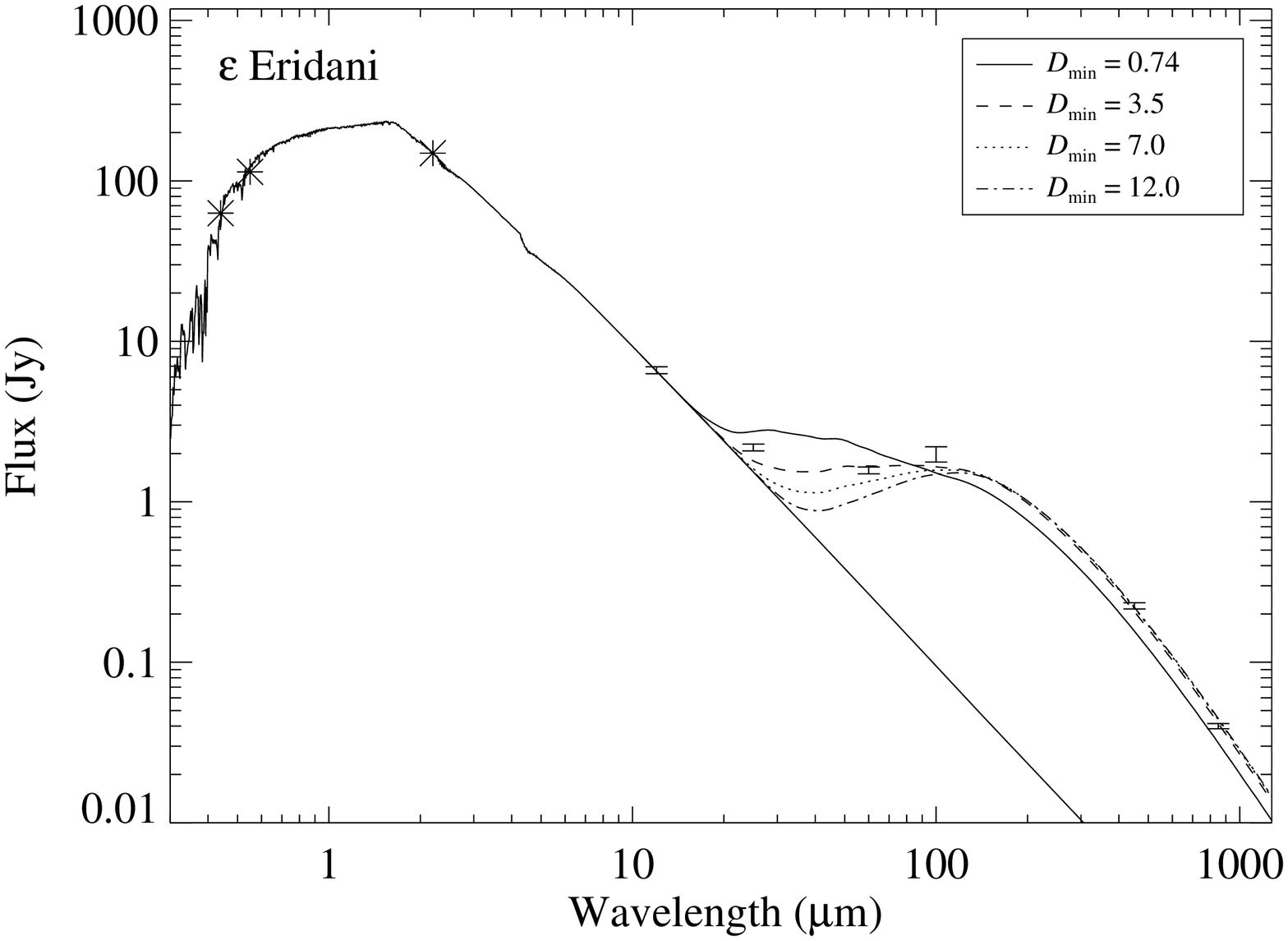,angle=90,width=\columnwidth}
  \end{center}
  \caption{SED of \epseri, with the ring radius set at 60
    AU. Models show the effect of changing the minimum size cutoff
    when using solid grains ($p=0$).
    Colour corrections for the \iras data were estimated from the
    $D_{\mathrm{min}}=3.5$~\um model.\label{fig:epseri_size}}
\end{figure}

\begin{figure}
  \begin{center}
    \psfig{figure=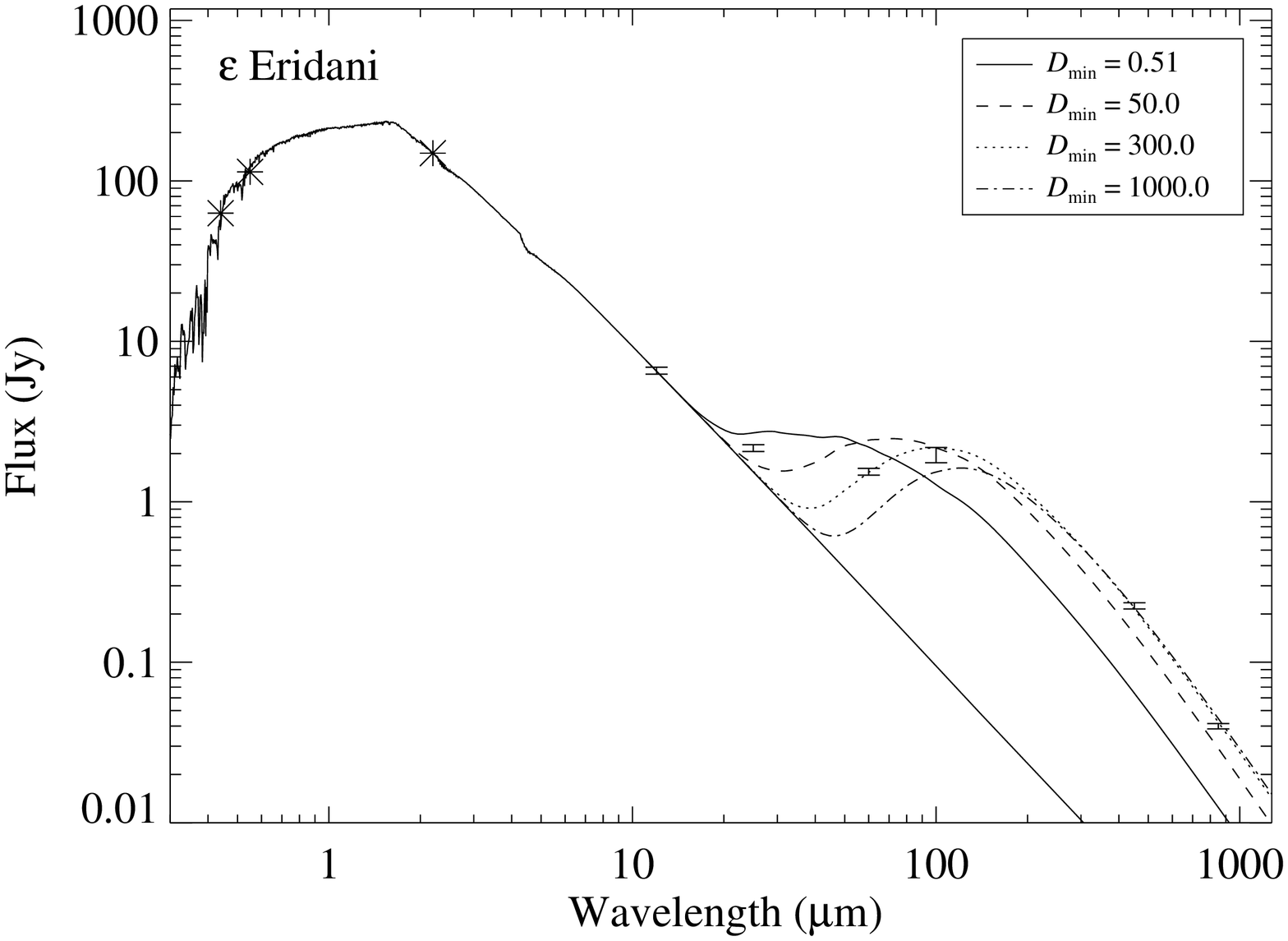,angle=90,width=\columnwidth}
  \end{center}
  \caption{SED of \epseri, with the ring radius set at 60 AU. Models
    show the effect of changing the minimum size cutoff when using
    porous grains ($p=0.9$). Colour corrections for the \iras data were
    estimated from the $D_{\mathrm{min}}=300$~\um
    model.\label{fig:epseri_size_porus}}
\end{figure}

\begin{figure}
  \begin{center}
    \psfig{figure=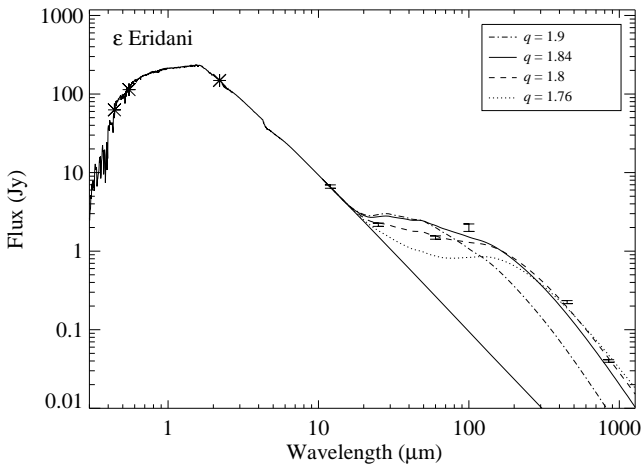,angle=90,width=\columnwidth}
  \end{center}
  \caption{SED of \epseri, with the ring radius set at 60
    AU. Models show the effect of changing the slope of the size
    distribution ($q$).
    Colour corrections for the \iras data were estimated from the
    $q=1.8$ \um model.\label{fig:epseri_slope}}
\end{figure}

In modelling \epseri's dust disk, a problem arises due to the low
luminosity of the star. Our model for the dust size distribution
assumes that radiation pressure will remove the smallest dust grains.
However, in the case of \epseri the radiation pressure is not
sufficient to blow dust grains out of the system when porous dust
grains are used (Figure~\ref{fig:epseri_beta}). In this situation, it
is not clear what limits the size of the smallest dust grains. As a
first step, we made the ad-hoc assumption that when porous grains were
used, the small size cutoff is set at the grain size where $\beta$ is
greatest, i.e.\ where the radiation force is most significant compared
to the gravitational force. Modelling using this assumption produced a
very interesting result: all of the models predict too much flux in
the mid-IR, and too little in the sub-mm
(Figure~\ref{fig:epseri_comp}).  This means that our models have too
many small (and therefore hot) dust grains. This is an unexpected
result, as there is no obvious mechanism for removing the small
grains. By varying the minimum size cutoff explicitly, an excellent
fit is possible if the cutoff size is set to 3.5~\um and the $p=0$
model is used, as shown in Figure~\ref{fig:epseri_size}. It is
also possible to obtain a good fit using porous grains, but only if
the minimum size cutoff is set to around 300~\um
(Figure~\ref{fig:epseri_size_porus}). This model does not predict the
observed excess at 25~\um, but this flux could come from an additional
hot component close to the star. Modifying the slope of the size
distribution can improve the fit (Figure~\ref{fig:epseri_slope}), but
cannot make the model consistent with the data.

Solid dust grains with a size of 3.5 \um have a $\beta$ of about 0.15,
much less than is needed to remove the grains by radiation
pressure. It therefore seems that the disk around \epseri contains few
grains smaller than about 3~\um, and radiation pressure from the star
cannot explain this deficiency.  Possible causes for this discrepancy
are discussed in Section~\ref{sec:discussion}.

Recently, detailed modelling of \epseri has been presented by
\nocite{2003ApJ...598L..51L}{Li}, {Lunine} \& {Bendo} (2003) who modelled the SED and SCUBA image using
a porous dust model ($p=0.9$, $q_{\mathrm{ice}}=0.0$), and were able
to get an excellent fit. With this in mind, our results seem somewhat
surprising, as they favour solid rather than porous grains, and fail
to produce an acceptable fit at all unless grains with diameter
smaller than 3.5~\um are excluded. The discrepancy arises because of
several differences in the assumptions made within the two models. The
most important difference is that \nocite{2003ApJ...598L..51L}{Li} {et~al.} (2003) set a
maximum grain size cutoff at a diameter of 2~cm, whereas in our
modelling the maximum grain size is set to be 10~m. If the collisional
cascade model is true, then the real size distribution will extend up
to very large bodies (i.e. kilometre scale and bigger), but the limit
of 10~m is chosen because bodies of this size and larger make a
negligible contribution to the sub-mm flux (i.e. the longest
wavelength data included in our fits), and so can be safely excluded
from the model. This is not true of 2~cm bodies, as grains of around
this size can make a very significant contribution the the flux at
850~\um. Similarly, \nocite{2003ApJ...598L..51L}{Li} {et~al.} (2003) set an a priori limit
on the minimum grain diameter at 2~\um; our finding that we must
remove grains smaller than 3.5~\um is therefore in reasonable
agreement with this (the remaining discrepancy is due to the fact that
\nocite{2003ApJ...598L..51L}{Li} {et~al.} (2003) allow the slope of the size distribution
to be a free parameter, whereas in the models where we modified
$D_\mathrm{min}$, $q$ was fixed to be $11/6$). Finally,
\nocite{2003ApJ...598L..51L}{Li} {et~al.} (2003) have a different spatial dust
distribution, with a peak at 55~AU and FWHM of 30~AU, compared with
the infinitely thin ring at 60~AU used in our model. The broad spatial
distribution has not been considered here as it would make our SED fit
worse. Because \nocite{2003ApJ...598L..51L}{Li} {et~al.} (2003) set tight limits on the
size distribution, the dust SED becomes quite sharply peaked, and a
broad spatial distribution is needed to make the dust SED wide enough
to fit all the points. Conversely, in our model the broad size
distribution means that the model SED is already {\it too} wide, so a
broad spatial distribution would only make things worse. Given the low
signal to noise of the SCUBA image, a direct measurement of the width
of the dust ring is not possible.

\nocite{2000MNRAS.314..702D}{Dent} {et~al.} (2000) also modeled the \epseri data, and were
able to fit the SED using a modified blackbody. The fitted parameters
indicated that the disk must contain large grains ($\sim30$ \um), but
this type of modelling does not constrain the composition or porosity
of the dust grains.

\subsection{Vega}

\begin{figure}
  \begin{center}
    \psfig{figure=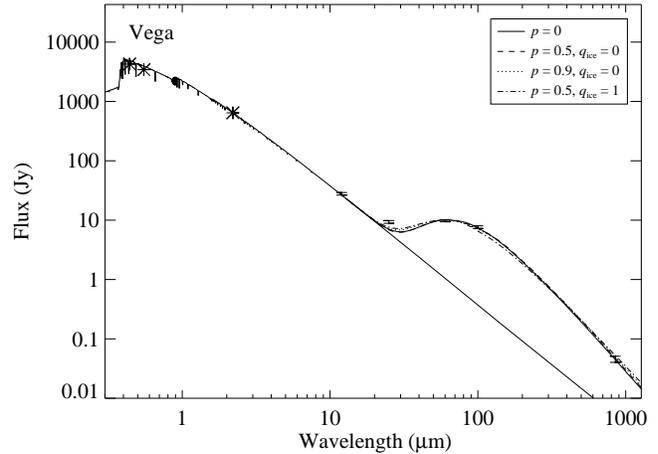,angle=90,width=\columnwidth}
  \end{center}
  \caption{SED of Vega and model fits to the data with ring radius as
    a free parameter. Colour corrections for the \iras data as plotted
    here were estimated from the $p=0$ model. \label{fig:vega_sed}}
\end{figure}

The original estimate for the size of Vega was made by
\nocite{1984ApJ...278L..23A}{Aumann} {et~al.} (1984), who analysed special pointed observations
made with \iras after the discovery of a far-IR excess. The estimated
diameter was 23~arcsec at 60~\um (corresponding to a radius of 90~AU).
Subsequently, this data was reanalysed, giving a larger diameter of
$35\pm5$ arcsec (135 AU) \nocite{1994A&A...285..229V}({van der Bliek}, {Prusti} \&  {Waters} 1994, and references
therein).  This discrepancy is attributed to the
fact the Aumann et al.\ did not account for the photospheric emission
at 60~\um, and hence underestimated the extent of the excess emission.
However, sub-mm observations made with SCUBA on the JCMT indicate a
diameter of $24\pm3$ arcsec \nocite{1998Natur.392..788H}({Holland} {et~al.} 1998), similar to
the original estimate from the \iras data. At millimetre wavelengths,
aperture synthesis imaging has been done using the Plateau de Bure
interferometer \nocite{2002ApJ...569L.115W}({Wilner} {et~al.} 2002) and Owens Valley Radio
Observatory \nocite{2001ApJ...560L.181K}({Koerner}, {Sargent} \&  {Ostroff} 2001). Both of these observations
suggested that the Vega disk is very clumpy, and found bright clumps at
a distance of 9~arcsec (Plateau de Bure) and 12~arcsec (Owens Valley)
from the central star. This could indicate a disk radius as small as
70 AU.

Given these observations, it is not obvious what the true size of the
disk is, and so we were unable to fix this parameter in our model
fits. Instead, we fitted different grain models with disk size as a
free parameter. With this approach we find that all of the dust grain
models fit equally well (Figure~\ref{fig:vega_sed}), but predict very
different disk sizes. The porous grain models both have a fitted
radius much larger than any of the observations, with the $p=0.5$,
$q_{\mathrm{ice}}=0.0$ model suggesting a radius of $191\pm16$ AU, and
the $p=0.9$, $q_{\mathrm{ice}}=0.0$ model giving $237\pm21$ AU. These
models can therefore be rejected. However, the solid grain model
($p=0.0$) gives a radius of $120\pm13$ AU, and the icy $p=0.5$,
$q_{\mathrm{ice}}=1.0$ model gives $92\pm8$ AU, both of which are
compatible with the observations. Without more detailed analysis of
the structure observed in the disk, it is impossible to distinguish
between these two options.

There is a small additional excess at 25~\um ($\sim 2$ Jy), which none
of the models account for. This could either be due to an additional
warm dust component similar to that found for HR4796, or because the
spatial distribution is broad, with dust at a range of distances from
the star.

\subsection{\betapic}

\begin{figure}
  \begin{center}
    \psfig{figure=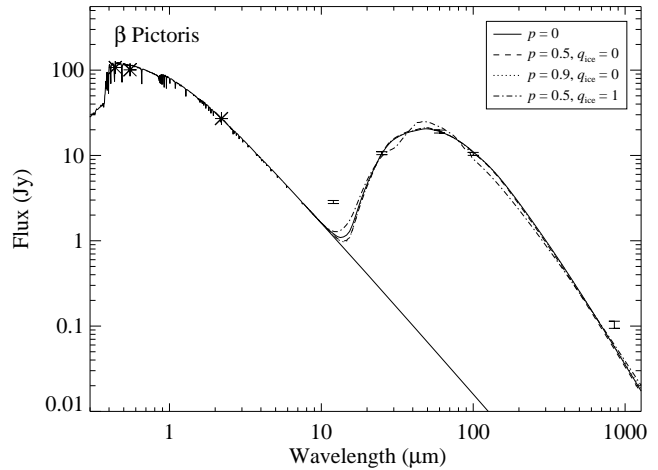,angle=90,width=\columnwidth}
  \end{center}
  \caption{SED of \betapic and model fits to the data with ring radius
    as a free parameter. Colour corrections for the \iras data as
    plotted here were estimated from the $p=0$ model.
    \label{fig:betapic_sed}}
\end{figure}

\betapic is the most extensively studied Vega excess star, with
observations in scattered light, mid-IR, far-IR, sub-mm. The system is
not a ring, but a disk with dust at a variety of distances from the
star. The system has been modelled in detail by
\nocite{1998A&A...331..291L}{Li} \& {Greenberg} (1998a), and our simpler model is not adequate to
deal with the extended spatial structure of the disk. However, we have
included \betapic here in order to show the consequences when our
model is applied to objects where the dust is not contained within a
ring. The ring morphology is one of the key assumptions of our model,
so it is extremely important that we can determine whether or not
this is true for unresolved objects.

To fit the SED, we used the same approach taken for Vega, i.e.\ we left
the ring radius as a free parameter, and fitted each dust composition
to the SED. The modelling shows that it is not possible to fit the
observed SED with any single model regardless of what dust composition
is chosen (Figure~\ref{fig:betapic_sed}). This is a helpful result, in
that it shows that we are able to distinguish between a disk and a
ring morphology purely from the SED. However, this modelling reveals
little else about the composition and size distribution of dust grains
in \betapic.

\subsection{HD141569}

HD141569 has a resolved circumstellar disk, first directly detected in
near-IR scattered light images \nocite{1999A&A...350L..51A,
  1999ApJ...525L..53W}({Augereau} {et~al.} 1999a; {Weinberger} {et~al.} 1999), and subsequently resolved in mid-IR thermal
emission \nocite{2000ApJ...532L.141F, 2001A&A...372L..61M,
  2002ApJ...573..425M}({Fisher} {et~al.} 2000; {Mouillet} {et~al.} 2001; {Marsh} {et~al.} 2002).  Most recently, the disk has been observed
with the ACS coronagraph \nocite{2002AAS...201.2501C}({Clampin} {et~al.} 2002), revealing that
the structure observed in the disk is probably caused by a tidal
interaction with a bound binary system at a projected distance of
around 1000~AU. The age of the system has been estimated to be $5\pm3$
Myr \nocite{2000ApJ...544..937W}({Weinberger} {et~al.} 2000), making it one of the youngest stars
in our sample. However, despite its youth the disk appears to be in a
collisional cascade, as modelling shows that it contains small dust
grains which must leave the system on a very short timescale
\nocite{2000ApJ...532L.141F}({Fisher} {et~al.} 2000).

\begin{figure}
  \begin{center}
    \psfig{figure=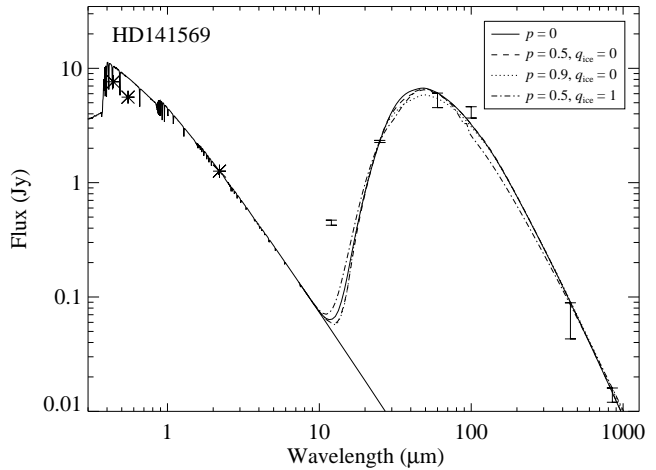,angle=90,width=\columnwidth}
  \end{center}
  \caption{SED of HD14159 with model fits for different grain
    compositions.  The ring radius is a free parameter in each case.
    Colour corrections for the \iras data as plotted here were
    estimated from the $p=0$ model. \label{fig:hd141569_sed}}
\end{figure}

The mid-IR imaging shows that the overall size of the disk is around
100 AU, but the emission is produced at a range of distances from the
star from 20~AU outwards \nocite{2002ApJ...573..425M}({Marsh} {et~al.} 2002). This is also
apparent from our SED modelling, as no model can simultaneously fit
all the data (Figure~\ref{fig:hd141569_sed}). When only the long
wavelength points are fitted (i.e.\ only the cooler, outer parts of the
disk), the fitted radius is about 50~AU if solid ($p=0$) or icy
($p=0.5$, $q_{\mathrm{ice}}=1.0$) grains are used, but 114~AU for
porous grains ($p=0.9$, $q_{\mathrm{ice}}=0.0$).  This indicates that
the thermal emission is probably produced by porous grains, in
agreement with the modelling done by \nocite{2003ApJ...594..987L}{Li} \& {Lunine} (2003b).

\begin{figure}
  \begin{center}
    \psfig{figure=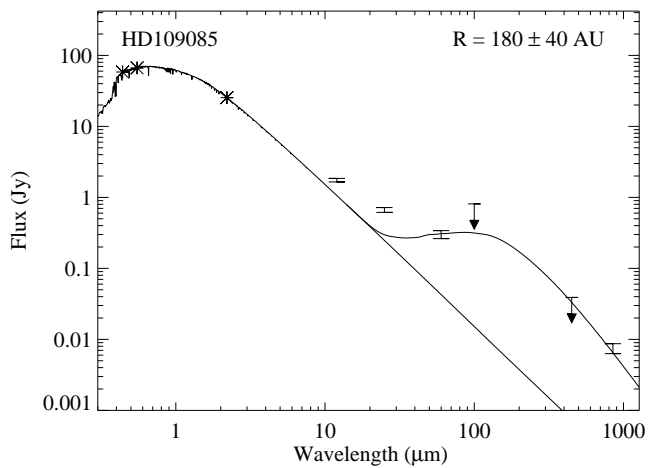,angle=90,width=\columnwidth}
  \end{center}
  \caption{SED of HD109085 with model fit to the data. Ring radius is
    a free parameter. Upper limits are 3~sigma. \label{fig:hd109085_sed}}
\end{figure}

\begin{figure}
  \begin{center}
    \psfig{figure=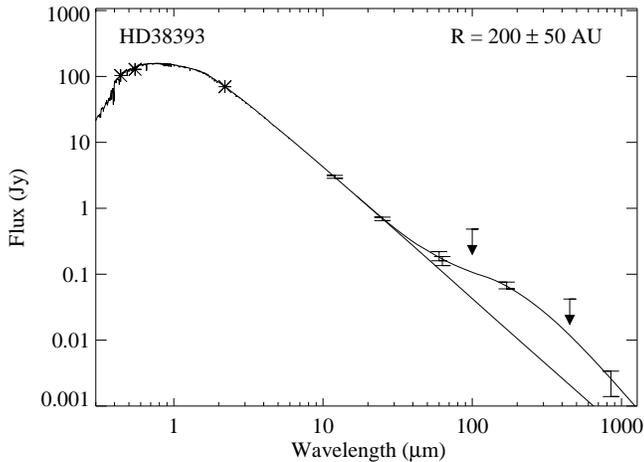,angle=90,width=\columnwidth}
    \end{center} \caption{SED of HD38393 with model fit to the
    data. The \iso 60~\um point has been offset to the right to
    distinguish it from the \iras 60~\um point. Upper limits are
    3~sigma. \label{fig:hd38393_sed}}
\end{figure}

\subsection{HD109085}

HD109085 is clearly detected in the sub-mm, indicating that the \iras
association is real. However, no model can simultaneously fit all of
the data (Figure~\ref{fig:hd109085_sed}). This indicates that there
must be dust at a variety of distances from the star, as is the case
for \betapic and HD141569. For our fitted model, the \iras 12~and
25~\um points have not been included, so the fitted radius reflects the
size of the coolest parts of the disk, i.e.\  the disk's overall
extent. This gives a result of 180~AU, which at a distance of 18.2 pc
gives an angular radius of 9.9 arcsec. This is more extended than the
SCUBA beam, so our flux estimate is probably smaller that the true
flux by a factor of $\sim 2$ (Table~\ref{tab:extphot}). Given the size
and brightness of the disk, it should be possible to fully map
HD109085 with SCUBA, and such observations are currently being
performed (Wyatt et al. in preparation).

\subsection{HD38393} \label{sec:hd38393}

HD38393 is extremely faint in the sub-mm, and is only detected at the
2.4 sigma level. The modelling produces a good fit to the SED
(Figure~\ref{fig:hd38393_sed}), but with a disk radius of $200 \pm 50$
AU for solid dust grains. At the distance of HD38393 (9.0 pc), this
corresponds to a radius of 22~arcsec. This would place most of the
flux outside the SCUBA beam (even though the extended photometry mode
was used), implying that the true flux at 850~\um is much higher than
the measured flux. In fact, if this radius is correct we would only
detect emission if the source is edge on. Revising our estimate of the
850 \um flux results in a still larger fitted radius (because a higher
sub-mm flux indicates a cooler disk), and so does not resolve this
problem. Nor does choosing a different dust composition, as porous
grains tend to be warmer than solid grains, and so also increase the
estimated disk size.

Given the low significance of the detection, it is possible that the
850~\um detection is either spurious, or caused by a background
galaxy.  Using source counts of galaxies detected in blank fields
\nocite{2002MNRAS.331..817S}(e.g.\ Figure~11 in {Scott} {et~al.} 2002), we find there is a
6~per cent probability that a 2.4 mJy background source will fall into
the beam of SCUBA in extended photometry mode (which is sensitive over
a 300~arcsec$^2$ area).  \nocite{2001A&A...365..545H}{Habing} {et~al.} (2001) also concluded
that their 170~\um detection is likely to be due to a background
galaxy, given the low flux and the galaxy number counts at that
wavelength.  However, an alternative explanation is that the estimate
of the disk size is wrong: the difficulty fitting the SED of \epseri
may indicate that our model is generally unreliable for cool stars,
and hence the true size of the disk could be smaller than 200~AU. This
would mean that the disk is deficient in small grains compared with
our model.

\subsection{\sigboo}

\begin{figure}
  \begin{center}
    \psfig{figure=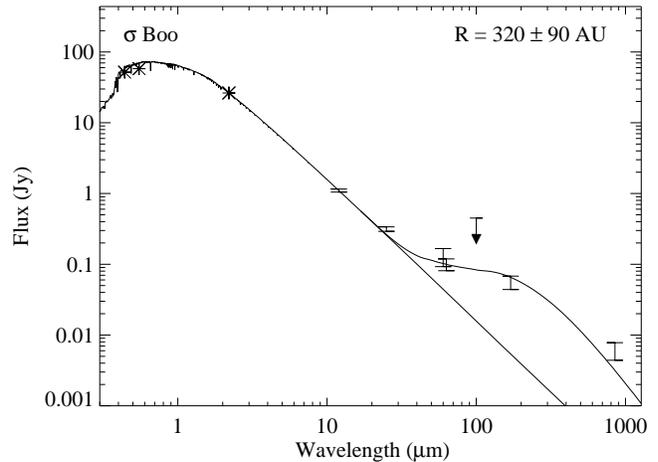,angle=90,width=\columnwidth}
  \end{center}
  \caption{SED of \sigboo. The \iso 60~\um point has been offset to the
    right to distinguish it from the \iras 60~\um point. Upper limits
    are 3~sigma. \label{fig:sigboo_sed}} 
\end{figure}

Excess sub-mm emission from \sigboo was detected at the 3.6 sigma
level, with a flux of 6.7 mJy. However, the best fitting model gives a
disk radius of $320\pm90$ AU, which corresponds to an angular radius
of 20~arcsec, implying that most of the 850~\um flux would be outside
the SCUBA beam even in extended photometry mode. In addition, the best
fitting model predicts a much lower 850~\um flux than the observed
value. As is the case for HD38393, there is no model which is
consistent with both the SED and the spatial constraints given by the
SCUBA beam size. The problem is unlikely to be caused by errors in the
far-IR data: the \iras and \iso results are entirely consistent, and
the \iso 170~\um observation has a beam radius of about 45~arcsec, so
all the flux would be detected with \iso, even if the disk really is
very large.

\nocite{2001A&A...365..545H}{Habing} {et~al.} (2001) considered the possibility that their \iso
detection at 60~and 170~\um was due to a background galaxy. They
concluded that the excess was probably real, but could not rule out
contamination by a background source. A background galaxy with a flux
of 6.7 mJy is unlikely to fall into the SCUBA beam by chance alone
(1 per cent probability), but as \sigboo was selected on the basis of a
far-IR excess, contamination by a background source becomes more
likely. The alternative explanation is that \sigboo is deficient in
small grains, which would reconcile the modelling results with the
SCUBA beam size.

\subsection{HD48682}

HD48682 is a quoted as visual binary in the Washington double star
catalogue with a separation of 34~arcsec. However the two stars cannot
be physically associated due to their differing proper motions, and
it appears that HD48682B is in fact a background object. Given that
the separation of the two components is similar to the \iras beam
size, it is not obvious whether the photosphere of the secondary will
contribute to the \iras fluxes or not. The \iras beam is different at
each wavelength, as the diffraction limit is much smaller at 12~\um
than it is at 100~\um. This means that the secondary photosphere may
only contribute to the measured fluxes at longer wavelengths. In
addition, it is very difficult to estimate the brightness of the
secondary at \iras wavelengths, because the star is most likely an M
star and only optical photometry is available. A small uncertainty in
the effective temperature therefore causes a large error in the
estimated mid-IR flux. A K band image would resolve this problem, but
unfortunately the secondary star is saturated on 2MASS images, so a
dedicated observation would be required. \nocite{1991ApJ...368..264A}{Aumann} \& {Probst} (1991)
made ground based observations of HD48682 at 10~\um, and concluded
that the 12~\um excess comes from $>6$ arcsec from the primary and is
consistent with the flux from the secondary, but that the secondary
photosphere cannot account for the large 60~\um excess. Our detection
with SCUBA confirms that the 60~\um \iras excess is associated with
the primary and not the secondary, but the source of the 12~and 25~\um
excess is not clear.

\begin{figure}
  \begin{center}
    \psfig{figure=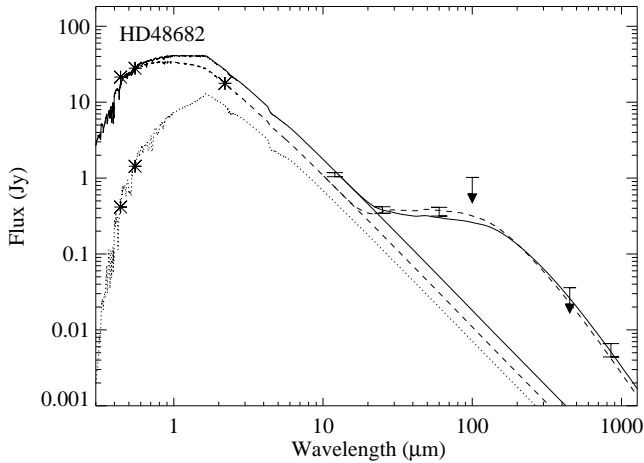,angle=90,width=\columnwidth}
  \end{center}
  \caption{SED of HD48682, with model fit to the data. Asterisks show
    optical and near-IR photometry, \iras and SCUBA data are shown with
    error bars. The photospheric flux from the primary, and the best
    fit model assuming that only the primary contributes to the \iras
    fluxes are shown with dashed lines. The secondary is shown with a
    dotted line. Solid lines show the combined photospheric flux, and
    combined photospheric flux plus best fit model assuming that the
    secondary does contribute to the \iras fluxes. Upper limits are
    $3\sigma$. \label{fig:hd48682_sed}}
\end{figure}

Figure~\ref{fig:hd48682_sed} shows the SED of HD48682, and the best
fit model. The modelling has been done in two ways, with and without
the secondary included in the photosphere subtraction (assuming a
spectral type M0). At 12~\um, the IRAS flux is more than the
photosphere of the primary, which could either be due to circumprimary
emission at $>6$ arcsec, or flux from the secondary. At 25~\um, the
\iras flux is more than the combined flux from primary and secondary,
indicating that there is definitely emission by dust at this
wavelength. There is a large excess at 60~\um, and the secondary
photosphere could only make a small contribution even if it is within
the beam. The modelling results are better if the secondary is
included in the photosphere subtraction (reduced $\chi^2=3.1$, verses
7.0 if the secondary is not included). If the secondary is not
included in the subtraction, the estimated radius is $71\pm15$ AU,
whereas if the secondary is included, the estimated radius is
$110\pm21$ AU. If the larger value is true, the projected radius would
be 6.7 arcsec, so the disk could be marginally resolved with SCUBA
mapping. The shape of the fitted model indicates that there would be
only a small excess at 18~\um, so resolved ground based imaging would
be difficult.

\subsection{HD207129}

\begin{figure}
  \begin{center}
    \psfig{figure=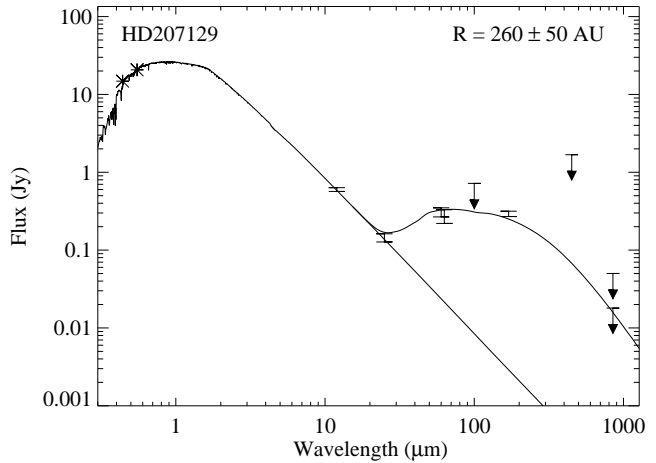,angle=90,width=\columnwidth}
  \end{center}
  \caption{SED of HD207129. The \iso 60~\um point has been offset to the
    right to distinguish it from the \iras 60~\um point. Upper limits
    are 3~sigma. The lower data point at 850~\um is the result if the
    source is unresolved, the upper point is the result if the source
    is resolved but with a radius smaller than 20~arcsec.
    \label{fig:hd207129_sed}}
\end{figure}

HD207129 is not detected in our reduction of archived SCUBA
measurements, but because there are \iso measurements at 170~\um a
model fit is possible, and the SCUBA upper limit also helps to
constrain this fit. The SED and model fit are shown in
Figure~\ref{fig:hd207129_sed}. The fitted size of the disk is
$260\pm50$ AU, which corresponds to an angular radius of
17~arcsec. Since this is larger than the SCUBA beam a significant
fraction of the flux would not have bean detected, and the quoted
upper limit on the 850~\um flux may actually be lower than the true
flux. HD207129 was observed in both photometry mode and map mode, and
the quoted upper limit of $<0.018$ Jy is based on both sets of
measurements. If just the mapping data is used, then an integrated
flux over a 20~arcsec radius aperture can be calculated. This method
gives a flux $<0.05$ Jy (3 sigma).

\section{Discussion} \label{sec:discussion}
The most striking result of our modelling is that there is a large
diversity in dust grain compositions between different disks. A simple
explanation for this result is that dust grain composition may vary
with age. The modelling of HD141569 and HR4796 (age 5~and 8~Myr
respectively) indicates high porosity grains ($p=0.9$), as does the
previous modelling of \betapic (age 20~Myr) done by
\nocite{1998A&A...331..291L}{Li} \& {Greenberg} (1998a). However, solid grains are indicated for
Fomalhaut and Vega (ages 150~and 350~Myr). It may be that the older
stars have less porous grains because the collisional cascade has
reprocessed the initially fluffy grains into a more solid form. An
alternative explanation is that the change in porosity is due to the
effects of stellar radiation on the dust, as suggested to explain the
differences between the porosity of long and short period comets
\nocite{1983P&SS...31..655M, 1998A&A...338..364L}({Mukai} \& {Fechtig} 1983; {Li} \& {Greenberg} 1998b). Clearly, a larger
sample of resolved disks is required to confirm this trend.

The failure of our basic model to fit the SED of \epseri is extremely
interesting, and it's important to determine what is causing the poor
fit. There are essentially two possibilities: either there is an
unknown physical effect which is modifying the size distribution in
the particular case of \epseri, or else the size distribution model we
have assumed is intrinsically wrong.

\epseri is known to have a planet \nocite{2000ApJ...544L.145H}({Hatzes} {et~al.} 2000), with a
semi-major axis of 3.4 AU and a mass of around 2~Jupiter masses
(assuming an inclination of $25^{\circ}$ as measured by
\nocite{1998ApJ...506L.133G}{Greaves} {et~al.}, 1998). This planet in unlikely to affect the
dust size distribution at 60~AU, because of its small orbital radius.
However, a planet with a large semi-major axis has been inferred from
the existence of clumps within the dust disk
\nocite{2002ApJ...578L.149Q}({Quillen} \& {Thorndike} 2002). If this planet is real, then dynamical
effects are clearly very important in shaping the disk, and this would
undoubtedly affect the size distribution of the dust (Wyatt 2003,
submitted).

The alternative explanation is that the discrepancy is not due to an
external influence, but that a more accurate treatment of the dust
grain size distribution in a collisional cascade is needed to account
for the SED. More detailed size distribution models allow the grain
strength to vary with grain size, and account for the the effects of
removing small grains on the overall size distribution
\nocite{1994P&SS...42.1079C, 1997Icar..130..140D}({Campo Bagatin} {et~al.} 1994; {Durda} \& {Dermott} 1997).  For \betapic,
extensive numerical models of this type have been produced, both for
the inner disk (Th{\' e}bault, Augereau \& Beust, submitted) and for
the outer disk \nocite{2000A&A...362.1127K}({Krivov}, {Mann} \&  {Krivova} 2000). Both of these studies show
significant deviation from the theoretical size distribution used in
this paper. This suggests that a more rigorous treatment of the
collisional cascade in \epseri would also give a significantly
different size distribution. However, it is not obvious that this
would produce a better fit to the SED (i.e.\ that the new size
distribution would have few grains smaller than 3~\um). In fact,
according to \nocite{1994P&SS...42.1079C}{Campo Bagatin} {et~al.} (1994) we should expect an
enhancement of grains just above the blowout radius as they are less
likely to be destroyed by collisions, which would make the problem
worse. Conversely, the modelling of \betapic's outer disk by
\nocite{2000A&A...362.1127K}{Krivov} {et~al.} (2000) suggests a reduction in the number of
grains just above the blowout limit, because they are destroyed by
small, fast moving grains as they are ejected from the system by
radiation pressure. However, as the stellar luminosity is much lower
for \epseri, this mechanism is likely to be less important. Only
detailed modelling specific to \epseri's dust disk can answer this
question.

The new sub-mm data significantly expands the sample of true Vega
excess stars detected at this wavelength. Regardless of
the dust properties, we are able to distinguish between a disk
morphology and a ring morphology. Out of the five new objects, only
HD109085 is disk like, indicating that this type is less common than
the ring like structure observed around Fomalhaut and \epseri.

One of the main aims of this work was to determine the size of an
unresolved disk from just its SED. However, the modelling of resolved
targets has shown that this may not be possible. The large diversity
in dust properties for the resolved disks mean that we cannot make a
reliable assumption as to the composition of the grains in an
unresolved disk. In addition, the poor fit to \epseri's SED shows that
the model may not in general contain all of the physics necessary to
account for the observations. Our estimates of the disk size are based
on assumption that the dust grains are solid (as for Fomalhaut), but
if the grain properties are different, then the size estimates will be
incorrect.

Future progress in this field depends on spatially resolving disks, as
this allows a direct measure of disk size and thus measurements of the
dust composition.  Our results indicate that there are probably a few
more targets that can be resolved using SCUBA, given sufficient
integration time. Also, ground based mid-IR observations also have the
ability to resolve some disks. However, large numbers of resolved
disks may have to wait until SIRTF and SOFIA become available.

\section{Conclusion}
We have presented new observations and modelling of Vega excess disks
with sub-mm data. Our observations expand the sample of Vega excess
stars detected in the sub-mm from 5 to~10, with a further 4~objects
which were not detected but have useful upper limits. We have fitted
the observed SEDs with models based on realistic dust grain
composition and size distribution. We find that dust grain composition
varies significantly between different objects, with younger disks
having more porous grains. For \epseri, our model fails to fit the
data unless there are less small grains in the system than expected
from a collisional cascade. This discrepancy may be due to an
inadequacy in our model of the size distribution, or the result of an
external influence such as a planet. For the unresolved targets, disk
size is estimated assuming solid dust grains, but given the observed
diversity in dust composition and the problems fitting \epseri, these
size estimates may not be reliable.

\vspace{7mm}

The James Clerk Maxwell Telescope is operated by the Joint
Astronomy Centre on behalf of the UK Particle Physics and Astronomy
Research Council. We wish to thank the staff at the JCMT for enabling
these observations, and an anonymous referee for their helpful and
positive comments.

\bsp

\bibliography{}

\label{lastpage}

\end{document}